\documentclass[twocolumn,prl,showpacs,preprintnumbers]{revtex4}
\usepackage{amssymb}
\usepackage{graphicx}
\usepackage{dcolumn}
\usepackage{bm}

\begin{document}

\title{Bose-Einstein condensation into non-equilibrium states studied by condensate
focusing}

\author{I. Shvarchuck$^{1}$}
\author{Ch. Buggle${}^{1}$}
\author{D.S. Petrov${}^{1,2}$}
\author{K. Dieckmann${}^{1}$}\altaffiliation[Presently at: ]{MIT, Cambridge, MA 02139, USA.}
\author{M. Zielonkovski${}^{1}$}\altaffiliation[Presently at: ]{SAPMarkets, 69190 Walldorf, Germany.}
\author{\\M. Kemmann${}^{1}$}
\author{T. Tiecke$^{1}$}
\author{W. von Klitzing${}^{1}$}
\author{G.V. Shlyapnikov${}^{1,2}$}
\author{J.T.M. Walraven${}^{1}$}
\affiliation{${}^1$ FOM Institute for Atomic and Molecular
Physics\mbox{,} Kruislaan 407, 1098 SJ Amsterdam, The
Netherlands\\${}^2$ Russian Research Center, Kurchatov Institute,
Kurchatov Square, 123182 Moscow, Russia }
\date{\today}

\begin{abstract}
We report the formation of Bose-Einstein condensates into
non-equilibrium states. Our condensates are much longer than
equilibrium condensates with the same number of atoms, show strong
phase fluctuations and have a dynamical evolution similar to that
of quadrupole shape oscillations of regular condensates. The
condensates emerge in elongated traps as the result of
\emph{local} thermalization when the nucleation time is short
compared to the axial oscillation time. We introduce condensate
focusing as a powerful method to extract the phase-coherence
length of Bose-Einstein condensates.
\end{abstract}

 \maketitle

Among the quantum fluids, the quantum gases are especially suited
to study the kinetics of Bose-Einstein condensation (BEC). The
modest density of the quantum gases makes it possible to stretch
the time of condensate formation to values allowing detailed
experimental investigation. In external potentials condensates
have a characteristic equilibrium shape, known as the Thomas-Fermi
shape, that differs markedly from the shape of thermal clouds.
Since the pioneering experiments on BEC in the alkali systems
\cite {Ander95,Davis95,Bradl95}, this feature has been observed by
many groups and is used routinely to measure the condensate
fraction \cite{Kette??}. Phase coherence is another key property
of equilibrium condensates and was established in the first
interference experiments \cite{Andre97,Steng99,Hagle99,Bloch00}.

Equilibrium condensates \cite{Ander95,Davis95} are produced by quasi-static
growth, where heat extraction limits the formation rate. The condensate
nucleates as a small feature in the center of the trap and grows as long as
heat is extracted from the sample. To observe the formation kinetics, the
gas has to be brought out of equilibrium, in practice by shock cooling.
Since the first experiment on condensate growth, by Miesner\thinspace \emph{%
et\thinspace al.\thinspace }\cite{Miesn98}, this is done by fast RF removal
of the most energetic atoms from the trap. Starting from a thermal gas just
above the phase transition temperature ($T_{C}$), the condensate appears as
the result of thermalization. Miesner\thinspace \emph{et\thinspace al.~}\cite
{Miesn98} observed the growth under adiabatic conditions. K\"{o}hl\thinspace
\emph{et\thinspace al.~}\cite{Kohl02} continued the extraction of heat and
atoms, also during growth. In both experiments, the condensate was observed
to grow from the center of the trap, like in the quasi-static limit.

Kagan\thinspace \emph{et\thinspace al.\thinspace }\cite{Kagan92}
pointed out that qualitatively different stages have to be
distinguished in the formation of equilibrium condensates with a
large number of atoms. The early stage (kinetic stage) is governed
by Boltzmann kinetic processes and leads to a preferential
occupation of the lowest energy levels. Once a substantial
fraction of the atoms gathers within an energy band of the order
of the chemical potential of the emerging condensate during
formation, their density fluctuations are suppressed in a fast
interaction-dominated regime governed by a non-linear equation for
the boson field. The appearing phase-fluctuating condensate then
grows and the condensed fraction approaches its equilibrium value.
However, the phase fluctuations still persist, giving rise to
dynamically evolving flow patterns in search for the true
equilibrium state. In elongated 3D trapped gases the phase
fluctuations can be pronounced even under equilibrium conditions
as was predicted by Petrov\thinspace \emph{et\thinspace
al.~}\cite{Petrov01} and observed experimentally by
Dettmer\thinspace \emph{et\thinspace al.~}\cite{Dettm01}.

In this Letter we report the formation of condensates into
non-equilibrium states and a new path towards equilibrium in
elongated traps. In contrast to the previous experiments our
results were obtained starting from thermal clouds deep in the
cross-over regime to hydrodynamic behavior. The condensates are
much longer than equilibrium condensates with the same number of
atoms. Moreover, they display strong phase fluctuations and a
dynamical evolution similar to that of a quadrupole shape
oscillation decaying towards equilibrium. We identify $1/\omega
_{z}$ as a characteristic time that should be addressed explicitly
for elongated cylindrical harmonic traps, i.e.~for traps with
$\omega _{\rho }\gg \omega _{z},$ where $\omega _{\rho }$ and
$\omega _{z}$ are the radial and axial angular frequencies,
respectively. We show that these exotic condensates emerge as the
result of \emph{local} thermalization when the nucleation time is
short as compared to $1/\omega _{z}$. The dynamical evolution of
the condensate in the trap has to be dealt with explicitly to
properly interpret time-of-flight absorption images. In this
context we introduce condensate focusing as an alternative to
Bragg scattering \cite{Steng99} for measuring the phase-coherence
length of phase-fluctuating Bose-Einstein condensates.

In the previous experiments on condensate formation the phase fluctuations
were not studied. The results of Miesner\thinspace \emph{et\thinspace
al.\thinspace }\cite{Miesn98} were compared to an analytical expression for
adiabatic growth of a condensate from a thermal cloud, derived by
Gardiner\thinspace \emph{et\thinspace al.\thinspace }\cite{Gardi97}.
Although a qualitative agreement between theory and experiment was readily
obtained it turned out to be impossible to obtain detailed agreement at the
quantitative level \cite{Gardi98,Gardi00}. In the experiment of
K\"{o}hl\thinspace \emph{et\thinspace al.\thinspace }\cite{Kohl02}
quantitative agreement with the quantum kinetic approach (see refs. \cite
{Gardi97,Gardi98,Gardi00}) was obtained for strong truncation, whereas for
weak truncation the observed behavior differed distinctly from theory.

In our experiments we typically load $4\times 10^{9}$ atoms of $^{87}$Rb in
the $|5^{2}S_{1/2},F=2,\ m_{F}=2\rangle $ state into a horizontal
Ioffe-Pritchard quadrupole trap with $\omega _{\rho }=2\pi \times 477(2)$ Hz
and $\omega _{z}=2\pi \times 20.8(1)$ Hz (no RF dressing) \cite{Dieck01}.
The trap minimum $B_{0}=88.6(1)$~$\mu $T corresponds to a radio frequency of
$\nu _{0}=620$ kHz as calibrated against atom laser output coupling \cite
{Bloch99}. The trap minimum shows a long term drift of 5~kHz/hr. At full
current thermal drift effects are less than 1~kHz/s.

The gas is prepared by forced RF evaporation at a final rate of $\dot{\nu}_{%
\text{tr}}=-433$ kHz/s to a value $\nu _{\text{tr,a}}=740$~kHz, followed by
20 ms of plain evaporation at $\nu _{\text{tr,a}}.$ As
\begin{equation}
|\dot{\nu}_{\text{tr}}/(\nu _{\text{tr,a}}-\nu _{0})|\ll \omega _{z}
\end{equation}
this yields a static, purely thermal cloud of $N_{i}\approx
5\times 10^{6}$ atoms at a temperature $T_{0}=1.3(1)~\mu $K as
determined from the axial size $l _{z}=[2kT_{0}/m\omega
_{z}^{2}]^{1/2}$ of the cloud in the trap, measured by time of
flight absorption imaging shortly $(t<1/\omega
_{z})$ after release from the trap. We calculate a central density $%
n_{0}\approx 4\times 10^{14}$~cm$^{-3}$, corresponding to a mean free path $%
\lambda _{0}=(2^{1/2}n_{0}\sigma )^{-1}\approx 3$~$\mu $m and a collision
rate $\tau _{\mathrm{col}}^{-1}\approx n_{0}v_{th}\sigma \approx 5000$~s$%
^{-1}$ \cite{Kemp02}. Radially we find $\lambda _{0}/l _{\rho
}\approx 0.5.$ Axially we have $\lambda _{0}/l _{z}\approx 0.02$
to be compared with the values $\lambda _{0}/l _{z}\approx 0.5$
and $0.3$ used in previous experiments \cite{Miesn98} and
\cite{Kohl02} on condensate formation. Hence, our thermal samples
are prepared far deeper in the cross-over to the hydrodynamic
regime.

The hydrodynamic behavior manifests itself in the damping time $\tau _{Q}$
and a shift of the frequency $\omega _{Q}$ of the quadrupole shape
oscillation as well as in an anisotropic expansion of the cloud after
release from the trap. We measured $\omega _{z}\tau _{Q}=10(3)$ and $\omega
_{Q}/\omega _{z}=1.56(5)$. For $\omega _{z}\tau _{Q}=10(3)$ theory predicts $%
\omega _{Q}/\omega _{z}=1.57(2)$ \cite{Guery99} and $\omega _{Q}/\omega _{z}=%
\sqrt{12/5}\approx 1.55$ for the hydrodynamic limit in very elongated traps
\cite{Kagan97,Griff97}. Further, writing $\beta \equiv \omega _{z}/\omega
_{\rho },$ from the scaling theory \cite{Kagan97}, we find in this limit $%
v_{z}/v_{\rho }\approx 2.6\beta \approx 0.11$ for the ratio of axial to
radial expansion velocity after release from the trap. We measured $%
v_{z}/v_{\rho }=0.78(2)$ and a final temperature $T_{\infty }=0.94(4)\ \mu $%
K as determined from the axial expansion, which implies that the gas expands
hydrodynamically and cools only briefly before the expansion becomes
ballistic. For isentropic hydrodynamic expansions the degeneracy parameter
is conserved: $n(\tau )/n_{0}=(T_{\infty }/T_{0})^{3/2}$. Then, assuming
that the expansion is hydrodynamic initially and becomes ballistic after a
time $\tau _{\mathrm{freeze}}$, the scaling theory \cite{Kagan97} gives a
relation between $T_{0}/T_{\infty }$ and $v_{z}/v_{\rho }$. In elongated
traps, for $v_{z}/v_{\rho }\gg\beta$, we find
\begin{equation}
T_{0}/T_{\infty }\approx [1+2(v_{\rho }/v_{z})^{2}]/3.
\end{equation}
The observed ratio $v_{z}/v_{\rho }$ and temperature $T_{0}$ lead to $%
T_{\infty }=0.91(7)$, which coincides with $T_{\infty }$ measured
from the
axial expansion. From the obtained value of $n(\tau _{\mathrm{freeze}%
})/n_{0} $ we calculate $\tau _{\mathrm{freeze}}\approx 0.3$ ms,
i.e.~a couple of collision times. This is consistent with the
ratio $\lambda _{0}/l _{\rho }\approx 0.5$ mentioned above and
confirms the picture that we are operating near the onset of
hydrodynamic behavior in the radial direction. For completeness we
verified that the expansion becomes isotropic, $v_{z}/v_{\rho
}=1.02(4),$ when the atom number is reduced by a factor of $30$.

Once the thermal cloud is prepared we distinguish three distinct stages.
\emph{First}, in the truncation stage, the radio frequency is set to the
value $\nu _{\text{tr,b}}=660$~kHz. This stage has a duration $t_{\text{tr}%
}=1$~ms, which is chosen to be long enough ($t_{\text{tr}}>1/\omega _{\rho }$%
) to allow atoms with radial energy $\varepsilon _{\rho }$ larger
than the RF truncation energy $\varepsilon _{\mathrm{tr}}$ to
escape from the trap, yet is short enough to disallow evaporative
cooling. We found that in this stage $50\%$ of the atoms are
removed. Notice that due to the finite radial escape horizon
$(\lambda _{0}/l _{\rho }\approx 0.4)$ the ejection is not
expected to be complete. Futhermore, the escape efficiency is
anisotropic as
a result of gravitational sag. The truncation energy $\varepsilon _{\mathrm{%
tr}}$ covers the range $3$ $\mu $K - $5$ $\mu $K depending on the position
of the truncation edge in the gravity field and is lowered by an additional $%
1$ $\mu $K due to RF-dressing (Rabi frequency $\Omega _{\text{rf}}\approx
2\pi \times 14$ kHz). At the start of the \emph{second} stage, the
thermalization stage, the radio frequency is stepped back up for a time $t_{%
\text{th}}$ to the frequency $\nu _{\text{tr,a}}$ to allow the gas to
thermalize under formation of a condensate. The value $\nu _{\text{tr,a}}$
is chosen to eliminate any appreciable evaporative cooling. The \emph{third}
stage, the expansion stage, starts by switching off the trap and covers the
time of flight $\tau $ after which the sample is absorption imaged on the $%
|5^{2}S_{1/2},F=2\rangle \rightarrow \,|5^{2}P_{3/2},F=1,2$ or $3\rangle $
transition.

To follow the evolution of the trapped gas after the truncation we took
time-of-flight absorption images for a range of evolution times $t\equiv t_{%
\text{tr}}+t_{\text{th}}$ and a fixed expansion time $\tau .$ The images
show a bimodal distribution, indicating that the truncation procedure
results in BEC. The condensate fraction grows to a final value of 6\% with a
characteristic time of 6 ms. This corresponds to $~30\tau _{\mathrm{col}},$
in accordance with previous experiments.

Rather than discussing the details of the growth kinetics we emphasize that
our condensates nucleate into non-equilibrium states. In Fig.1 we plot the
Thomas-Fermi half-length $L_{z}$ obtained with the standard fitting
procedure of a bimodal distribution to our data \cite{Kette??}. For the
shortest expansion time, $\tau =2.8$~ms, the axial size of the condensate
image equals to good approximation $(\tau \ll 1/\omega _{z})$ the axial size
of the condensate in the trap. We see that $L_{z}(t)$ is initially \emph{%
oversized} by a factor $L_{z}(0)/L_{z}(\infty )=2.2(3)$ and rapidly
decreases to reach its equilibrium size after roughly one strongly damped
shape oscillation (see open triangles Fig.1). Hence, the condensate is
clearly not in equilibrium \cite{Note1}.

\begin{figure}[tbp]
\includegraphics[width=83mm]{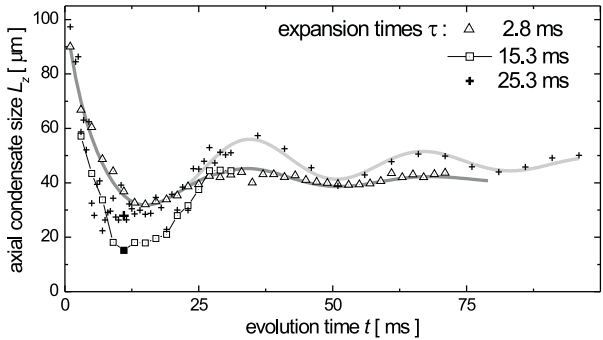}% Here is how to import EPS art
\caption{Condensate length $L_{z}(t,\protect\tau )$ versus evolution time $t$
for three different expansion times $\protect\tau $. The dark grey line is a
guide to the eye. The light grey line represents a fit to a damped
quadrupole shape oscillation.}
\label{fig:1}
\end{figure}

The formation of oversized condensates follows from the local formation
concept underlying ref.~\cite{Kagan92}. Starting with a thermal cloud of $%
N_{i}=5\times 10^{6}$ atoms in an $11.5$ mK deep harmonic trap at
a temperature $T_{0}=1.3~\mu $K, we calculate that $55\%$ of the
atoms remain
trapped after all atoms with energy $\varepsilon >\varepsilon _{\mathrm{tr}%
}=3.4$~$\mu $K are removed. The gas will rethermalize well within a time
short compared to $1/\omega _{z}$. Hence, the resulting temperature varies
along the trap axis. In this respect the thermalization is a \emph{local}
phenomenon. The local $T_{C}$ is given by \cite{Pethic02}
\begin{equation}
kT_{C}(z)\approx 1.28~\hslash \omega _{\rho }\left[ n_{\text{1D}}(z)r_{\rho }%
\right] ^{2/5},
\end{equation}
where $r_{\rho }=[\hslash /m\omega _{\rho }]^{1/2}$ is the radial
oscillator
length and $n_{\text{1D}}(z)$ the atom number per unit length at position $z$%
. We find that the local temperature $T(z)$ is lower than the
local $T_{C}(z) $ over a length of order $l _{z}$. In view of the
simplicity of this model we consider this as good qualitative
agreement with experiment.

To further investigate the formation process we introduce condensate
focusing. In our case one-dimensional focusing results from axial
contraction of the expanding cloud when the gas is released from the trap
during the inward phase of a shape oscillation. The focus is best
demonstrated by plotting the axial size $L_{z}(t,\tau )$ of the condensate
as a function of expansion time $\tau $ after a fixed evolution of $t=11$~ms
in the trap (see Fig.2). The axial size is seen first to decrease and to
increase again later as expected for a focus \cite{Note2}.

\begin{figure}[tbp]
\includegraphics[width=83mm]{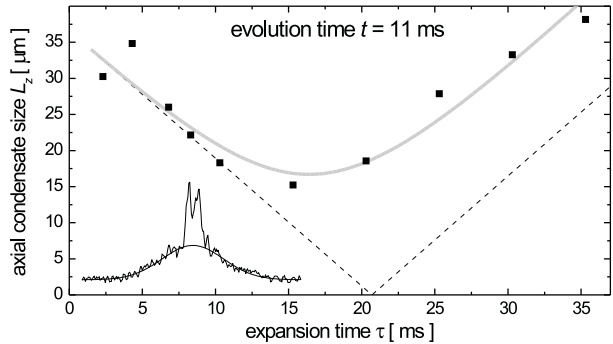}
\caption{Condensate length $L_{z}(t,\protect\tau )$ versus expansion time$~%
\protect\tau $ after 11~ms of evolution (10 point averages). The grey
(dashed) line is a fit to the scaling equations including (excluding) phase
fluctuation broadening. Inset: axial density profile showing phase
fluctuation stripes at $\protect\tau = 6.8$ ms. }
\label{fig:2}
\end{figure}

We elucidate the focusing in relation to the low-frequency branch of the $%
m=0 $ quadrupole shape oscillation of the Thomas-Fermi condensate.
In the limit of linear response the scaling parameter $b_{z}$ for
the axial size of the condensate in the trap can be written as
$b_{z}(t)\equiv L_{z}(t)/L_{z}(0)=1+a_{z}\cos \omega _{Q}t, $ with
$a_{z}\ll 1$ the scaled axial amplitude and $t$ the evolution
time. Releasing the gas at time $t$ and observing the cloud after
an expansion time $\tau \gg 1/\omega _{\rho},$ the scaling theory
\cite{Castin96,Kagan97} offers an approximate expression for the
scaled axial size
\begin{equation}
b_{z}(t,\tau )\approx 1+(\pi \beta \omega _{z}/2-a_{z}\omega _{Q}\sin \omega
_{Q}t)\tau .  \label{eq3}
\end{equation}
The first term under the brackets corresponds to the axial expansion kick
caused by the declining chemical potential at trap release. The second term
represents the scaled axial dilatation velocity in the trap at the moment of
release. From Eq.(\ref{eq3}) we see that the axial dilatation field reaches
a real (virtual) focus for positive (negative) values of the expansion time $%
\tau _{\text{focus}}= (a_{z}\omega _{Q}\sin \omega _{Q}t-\frac{\pi
}{2}\beta \omega _{z})^{-1}.$ A real focus can be obtained already
for small amplitudes, $a_{z}>\beta $.

For $\tau \gg 1/\omega _{\rho }$ the radial expansion is described
by $ b_{\rho }(t,\tau )\approx \omega _{\rho }\tau (1-\textstyle{\frac{1}{4}}%
a_{z}\cos \omega _{Q}t)$, i.e. shows no focus. As the radial size
remains finite, the chemical potential will build up near the
focus until the compression is balanced and the axial size starts
to expand again. As follows from the scaling equations (see
ref.~\cite {Castin96} or \cite{Kagan97}), at maximum compression
the axial size is reduced by a factor of $\beta ^{2}.$

Our data match a scaled focal size $b_{z}(t,\tau
_{\text{focus}})=0.49(6),$ i.e.~the focus is strongly broadened as
compared to the minimum scaled size $\beta ^{2}$. The broadening
is attributed to local variations in expansion velocity caused by
the presence of phase fluctuations in our condensates. After some
expansion these variations give rise to irregular stripes (see the
inset of Fig.2) as previously observed in Hannover \cite{Dettm01}.
At the focus, the axial distribution maps linearly onto the
momentum distribution due to the phase fluctuations in the
original
condensate. The scaled focal size is given by $b_{z}(t,\tau _{\text{focus}%
})\sim (\hbar/mL_{\phi})\tau _{\text{focus}}/L_{z}$%
, where $L_{\phi}$ is the phase coherence length and
($\hbar/mL_{\phi}$) characterizes the expansion velocity due to
the phase fluctuations. For equilibrium phase fluctuations
close to $T_{C}$ we estimate $%
L_{z}/L_{\phi }=7(4)$ (see ref.~\cite{Note3}) and $b_{z}\sim
0.05$. As the observed focal size is larger by an order of
magnitude, the phase fluctuations have a non-equilibrium origin
with a phase coherence length of only $L_{\phi }\sim 1$~$\mu $m.
The decay of the phase fluctuations as a function of evolution
time is subject of further investigation in our group.

We point out that for a \emph{thermal} cloud driven on the
low-frequency quadrupole mode the phenomenon of focusing is absent
except deeply in the hydrodynamic regime. For a small ratio
$T_{\infty }/T_{0}$ the scaled focal size will be given by
$b_{z}(t,\tau _{\text{focus}})\approx v_{th}(T_{\infty})\tau
_{\text{focus}}/l_{z}(T_{0})\approx (T_{\infty
}/T_{0})^{1/2}\omega _{z}\tau _{\text{focus}}<1$.

In Fig.1 we also show the oscillation in the axial size of the condensate as
observed for $15.3$~ms (open squares) and $25.3$~ms (crosses) of expansion
\cite{Note4}. Due to enhancement by focusing the amplitude of the
oscillation has increased as compared to the $2.8~$ms results. For $\tau
=25.3$~ms the shape oscillation is seen to exceed the noise for at least $100
$~ms. This oscillation can be described by a linear response expression for
evolution times $t\geq 20$~ms where $a_{z}\leq 0.2$. We measure a damping
time of $\tau _{Q}=50(9)$~ms and a frequency ratio $\omega _{Q}/\omega
_{z}=1.54(4)$. The latter is slightly lower than the frequency expected for
a quadrupole shape oscillation of a pure condensate in very elongated traps,
$\omega _{Q}/\omega _{z}\approx \sqrt{5/2}\approx 1.58$ \cite{Kagan97}\cite
{Castin96}. A 5\% negative frequency shift was observed for the quadrupole
mode in Na condensates just below $T_{C}$ \cite{Stamp98} and is consistent
with theory \cite{Dalvo99}.

The condensate focusing is a particularly useful tool. It
separates the condensate from the thermal cloud in time-of-flight
absorption imaging. This enables the observation of small
condensate fractions and measurements close to $T_{C}.$ It also
allows the observation of small thermal clouds and thus extending
of the dynamic range of time-of-flight thermometry. The focal size
and shape allow to quantify the phase fluctuations inside
condensates. Extended to two dimensions the focusing can serve
imaging applications, with the mean field of the quantum fluid
acting as a tunable component in atom optics.

This work is part of the Cold Atoms program of FOM and financially supported
by NWO under project 047.009.010, by INTAS under project 2001.2344, and by
the Russian Foundation for Basic Research (RFBR).


\begin{thebibliography}{99}
\bibitem{Ander95}  M. H. Anderson, J. R. Ensher, M. R. Matthews, C. E.
Wieman, E. A. Cornell, Science \textbf{269}, 198 (1995).

\bibitem{Davis95}  K. B. Davis, M. -O. Mewes, M. R. Andrews, N. J. van
Druten, D. S. Durfee, D. M. Kurn, and W. Ketterle, Phys. Rev. Lett. \textbf{%
75}, 3969 (1995).

\bibitem{Bradl95}  C.C. Bradley, C.A. Sackett, J.J. Tollett, and R.G. Hulet,
Phys. Rev. Lett. \textbf{75}, 1687 (1995);\emph{\ ibid} \textbf{78}, 985
(1997).

\bibitem{Kette??}  For a detailed description see W. Ketterle, D.S. Durfee
and D.M. Stamper-Kurn, in Proc. Int. School Phys. \textit{Enrico Fermi}
course CXL, M. Inguscio, S. Stringari, C. Wieman (Eds.), IOS Press,
Amsterdam (1999).

\bibitem{Andre97}  M.R. Andrews, C.G. Townsend, H.-J. Miesner, D.S. Durfee,
D.M. Kurn, W. Ketterle, Science, \textbf{275}, 637 (1997).

\bibitem{Steng99} J. Stenger \emph{et\thinspace al.}, Phys. Rev. Lett.
\textbf{82} 4569 (1999).

\bibitem{Hagle99}  E.W. Hagley \emph{et\thinspace al.}, Phys. Rev. Lett.
\textbf{83}, 3112 (1999).

\bibitem{Bloch00}  I. Bloch, T.W. H\"{a}nsch and T. Esslinger, Nature
\textbf{403}, 166 (2000).

\bibitem{Miesn98}  H.-J. Miesner, D.M. Stamper-Kurn, M.R. Andrews, D.S.
Durfee, S. Inouye, and W. Ketterle, Science \textbf{279}, 1005 (1998).

\bibitem{Kohl02}  M. K\"{o}hl, M.J. Davis, C.W. Gardiner, T.W. H\"{a}nsch
and T. Esslinger, Phys. Rev. Let. \textbf{88}, 80402 (2002).

\bibitem{Kagan92}  Yu.M. Kagan, B.V. Svistunov, and G.V. Shlyapnikov, Sov.
Phys. JETP \textbf{75}, 387 (1992).

\bibitem{Petrov01}  D.S. Petrov, G.V. Shlyapnikov, and J.T.M. Walraven,
Phys. Rev. Lett. 87, 50404 (2001).

\bibitem{Dettm01}  S. Dettmer \emph{et\thinspace al.}, Phys. Rev. Lett. 87,
160406 (2001).

\bibitem{Gardi97}  C.W. Gardiner, P. Zoller, R.J. Ballagh, and M.J. Davis,
Phys. Rev. Lett. \textbf{79}, 1793 (1997).

\bibitem{Gardi98}  C.W. Gardiner, M.D. Lee, R.J. Ballagh, M.J. Davis, and P.
Zoller, Phys. Rev. Lett. \textbf{81}, 5266 (1998).

\bibitem{Gardi00}  M.J. Davis, C.W. Gardiner, and R.J. Ballagh, Phys. Rev. A
\textbf{62}, 63608 (2000); M.D. Lee and C.W. Gardiner, \emph{ibid} 33606
(2000); M.J. Bijlsma, E. Zaremba, and H.T.C. Stoof, \emph{ibid} 63609 (2000).

\bibitem{Dieck01}  For a detailed description of our apparatus see K.
Dieckmann, thesis, University of Amsterdam (2001).

\bibitem{Bloch99}  I. Bloch, T.W. H\"{a}nsch and T. Esslinger, Phys. Rev.
Let. \textbf{82}, 3008 (1999).

\bibitem{Kemp02}  Here $v_{th}=[8k_{\mathrm{B}}T_{0}/\pi m]^{1/2}$ is the
thermal velocity and $\sigma =8\pi a^{2}$ the elastic scattering
cross-section with $a=5.238(1)$~nm the s-wave scattering length; see E.G.M.
van Kempen, S.J.J.M.F. Kokkelmans, D.J. Heinzen, and B.J. Verhaar, Phys.
Rev. Lett. 88, 93201 (2002).

\bibitem{Guery99}  D. Gu\'{e}ry-Odelin, F. Zambelli, J. Dalibard, and S.
Stringari, Phys. Rev. A, \textbf{60}, 4851 (1999).

\bibitem{Kagan97}  Yu. Kagan, E.L. Surkov, and G.V. Shlyapnikov, Phys. Rev.
A \textbf{55}, R18 (1997).

\bibitem{Griff97}  A. Griffin, W.-C. Wu, and S. Stringari, Phys. Rev Lett.
\textbf{78}, 1838 (1997).

\bibitem{Note1}  The shock-cooling by truncation in the presence of
gravitational sag also gives rise to a vertical center of mass oscillation
of the sample in the trap with an amplitude of $0.5$ $\mu $m and a damping
time of 40 ms.

\bibitem{Pethic02}  C.J. Pethick and H. Smith, \emph{Bose-Einstein
condensation in dilute gases}, CUP, Cambridge (2002). See Eq.(2.19) with $%
\alpha =5/2$ and $N/L=n_{\text{1D}}(z)$, $L$ is the 1D volume.

\bibitem{Castin96}  Y. Castin and R. Dum, Phys. Rev. Lett. \textbf{77}, 5315
(1996).

\bibitem{Note2}  We observe the sample under $73^{\circ }$
with respect to symmetry axis of our trap. The data in Fig.2 are
corrected for this effect. This limits the observation of phase
stripes at long expansion times.

\bibitem{Note3}  From ref.\cite{Petrov01} we have for temperatures
close to $T_{C}$: $L_{z}/L_{\phi }\approx \delta
_{L}^{2}=(N/N_{0})(L_{z}/r_{z})^{2}(\beta N)^{-2/3}$, where $r_z$
is the axial harmonic oscillator length.

\bibitem{Note4}  In radial direction the oscillation amplitude did not
exceed the noise for the conditions of Fig.1. We note that giant shape
oscillations (also visible in the radial direction) were observed for a fast
RF ramp-down deep into the Bose-condensed regime, i.e. for $|\dot{\nu}_{%
\text{tr}}/(\nu _{\text{tr,a}}-\nu _{0})|>\omega _{z}$.

\bibitem{Stamp98}  D.M. Stamper-Kurn, H.-J. Miesner, S. Inouye, M.R.
Andrews, and W. Ketterle, Phys. Rev. Lett. \textbf{81}, 500 (1998).

\bibitem{Dalvo99}  See the review by F. Dalfovo, S. Giorgini, L.P.
Pitaevskii, and S. Stringari, Rev. Mod. Phys., \textbf{71}, 463 (1999).
\end{thebibliography}
\end{document}